\documentclass[11pt, letterpaper]{article}

\usepackage[utf8]{inputenc}
\usepackage[T1]{fontenc}
\usepackage{geometry}
\usepackage{amsmath, amsfonts, amssymb} 
\usepackage{graphicx}                   
\usepackage{booktabs}                   
\usepackage{hyperref}                   
\usepackage{float}                      
\usepackage{titlesec}                   
\usepackage{authblk}                    
\usepackage{enumitem}                   
\usepackage{caption}                    

\usepackage[none]{hyphenat}   
\usepackage{microtype}        
\hypersetup{colorlinks=true, linkcolor=blue, citecolor=blue, urlcolor=blue}
\title{\textbf{Citation-Enforced RAG for Fiscal Document Intelligence: Cited, Explainable Knowledge Retrieval in Tax Compliance}}

\author{Akhil Chandra Shanivendra\thanks{This work was conducted independently and does not represent the views of any employer or institution.}}
\affil{Independent Researcher \\ \textit{akhil.shanivendra@gmail.com}}

\date{}

\begin{document}

\maketitle


\begin{abstract}
\noindent Tax authorities and public-sector financial agencies rely on large volumes of unstructured and semi-structured fiscal documents—including tax forms, instructions, publications, and jurisdiction-specific guidance—to support compliance analysis and audit workflows. While recent advances in generative AI and retrieval-augmented generation (RAG) have shown promise for document-centric question answering \cite{Lewis2021}, existing approaches often lack the transparency, citation fidelity, and conservative behaviour required in high-stakes regulatory domains. In particular, unconstrained generation and summary-based multimodal retrieval can lead to hallucinated or weakly supported outputs that are difficult to verify. This paper presents a multimodal, citation-enforced RAG framework for fiscal document intelligence that prioritises explainability and auditability. The proposed system adopts a source-first ingestion strategy, embedding only content extracted directly from authoritative fiscal documents while preserving document identifiers and page-level provenance. Answer generation is constrained by explicit citation enforcement, which requires every generated claim to be grounded in retrieved evidence, and by an abstention mechanism that allows the system to decline to answer when document support is insufficient. The framework is evaluated on a corpus of real-world, publicly available tax documents from the U.S. Internal Revenue Service and two U.S. state tax authorities. Using a combination of automatic metrics and human evaluation, the study assesses retrieval performance, citation correctness, hallucination rates, abstention behaviour, and answer usefulness. Results demonstrate that the proposed design substantially improves citation fidelity and reduces unsupported claims while maintaining practical utility for analyst-facing tasks. Overall, this work illustrates how generative AI can be responsibly applied to multimodal fiscal document analysis by combining retrieval-based grounding with explicit constraints on generation, offering a practical pathway toward trustworthy AI systems for tax compliance and public-sector governance.
\end{abstract}

\section{Introduction}

\subsection{Background and Motivation}
Tax authorities and public-sector financial agencies operate in an environment characterised by an ever-growing volume of complex, unstructured documentation. Core compliance and audit workflows rely on a diverse set of fiscal artefacts, including scanned tax forms, filing instructions, policy manuals, audit memoranda, administrative notices, and jurisdiction-specific guidance documents. These materials are typically distributed as heterogeneous PDF files containing a mixture of free-form text, tables, structured layouts, and embedded images. As regulations evolve and reporting requirements change, maintaining accurate, consistent, and explainable access to this information presents a persistent operational challenge.

Traditional approaches to fiscal document analysis—such as rule-based systems, keyword search, and manually curated knowledge bases—struggle to scale in this setting. Rule-based pipelines require continuous maintenance and are brittle to changes in document structure or language. Keyword-based information retrieval systems, such as those relying on BM25 or Okapi algorithms, often fail to capture semantic intent, returning incomplete or contextually irrelevant results \cite{Robertson1995}. More recently, large language models (LLMs) have demonstrated impressive capabilities in natural language understanding and generation, motivating interest in their application to document-intensive public-sector workflows. However, naïvely applying LLMs to tax and compliance tasks introduces significant risks, particularly hallucinated responses \cite{Zhang2025}, unverifiable claims, and the absence of explicit links between generated answers and authoritative source documents.

In regulated domains such as taxation, explainability and traceability are not optional features but foundational requirements. Any automated system used to support compliance analysis or audit decision-making must allow analysts to verify outputs against the source materials, understand the provenance of each claim, and identify cases in which the available evidence is insufficient to support a definitive answer. These constraints sharply differentiate fiscal document intelligence from many consumer-facing or exploratory AI applications.

\subsection{Problem Statement}
Despite rapid progress in generative AI and retrieval-augmented generation (RAG) systems \cite{Lewis2021, IBM}, existing approaches remain poorly suited to high-stakes fiscal and tax compliance settings. Standard RAG pipelines typically focus on maximising answer fluency and semantic relevance, often embedding unstructured text or LLM-generated summaries into vector databases and relying on unconstrained generation at inference time \cite{Gao2023Survey}. While effective for general question answering, such designs offer limited guarantees regarding citation fidelity, evidence completeness, or control of hallucinations.

Multimodal RAG systems extend these ideas to documents containing images and tables, often by generating textual summaries of non-textual content and embedding them for retrieval. Although this strategy simplifies multimodal integration, it introduces an additional generative step before retrieval, effectively treating model-generated abstractions as primary evidence. In domains requiring auditability, this design choice obscures the relationship between system outputs and original documents, making it difficult to verify claims at the level of document spans, pages, or sections.

Furthermore, most existing systems assume that an answer should always be produced, even when retrieved evidence is weak or ambiguous. In tax compliance contexts, confidently stating an incorrect interpretation can be more harmful than explicitly acknowledging uncertainty. The lack of abstention mechanisms in current RAG systems, therefore, represents a critical gap for responsible deployment in public-sector analytics.

The central problem addressed in this work is the absence of a document-grounded, citation-enforced, and abstention-aware framework for multimodal fiscal document intelligence that satisfies the transparency and accountability requirements of tax compliance workflows.

\subsection{Contributions}
In this paper, this study presents a Generative AI and Multimodal Retrieval-Augmented Generation framework designed specifically for fiscal and tax-related document analysis under high accountability constraints. The contributions are as follows:

\begin{itemize}
    \item \textbf{Source-first multimodal RAG architecture:} The proposed architecture preserves document provenance by embedding and retrieving only extracted source content, rather than LLM-generated summaries, enabling span-level traceability.
    \item \textbf{Citation-enforced answer generation:} This study introduces a mechanism in which every generated statement must be explicitly grounded in retrieved evidence, with citations linked to document identifiers and page ranges.
    \item \textbf{Abstention-aware decision policy:} The framework incorporates a decision policy that allows the system to withhold answers when retrieved evidence is insufficient or unreliable, reducing hallucination risk in high-stakes scenarios.
    \item \textbf{Real-world evaluation:} The framework is evaluated on a corpus of real-world, publicly available fiscal documents from the U.S. Internal Revenue Service and two U.S. state tax authorities, without the use of synthetic or simulated data.
    \item \textbf{Comprehensive assessment:} Both automatic and human evaluations are conducted to assess retrieval quality, citation correctness, hallucination rates, abstention behaviour, and answer usefulness.
\end{itemize}

By situating generative AI within a rigorously constrained retrieval-and-citation framework, this work demonstrates how multimodal RAG systems can be adapted to support responsible, explainable decision-making in public-sector tax compliance.

\section{Related Work}
This work intersects several active research areas, including document intelligence for public-sector applications, retrieval-augmented generation (RAG), multimodal RAG systems, and explainable or trustworthy AI. This section reviews these strands of literature and highlights the limitations that motivate the proposed approach.

\subsection{Document Intelligence in Public-Sector and Financial Domains}
Automated document processing has long been a focus of research and practice in government and financial institutions. Early systems relied heavily on optical character recognition (OCR) pipelines combined with rule-based extraction and template matching to process structured forms and reports. In tax administration, such systems are commonly used to extract fields from standardised filings or to perform basic document classification. While effective for narrowly defined tasks, these approaches are brittle when faced with heterogeneous layouts, evolving regulations, and narrative guidance documents that do not conform to fixed templates. 

More recent work has explored machine learning–based document understanding, including layout-aware models such as LayoutLM \cite{xu2020layoutlm}, LayoutLMv3 \cite{huang2022layoutlmv3}, and DocFormer \cite{appalaraju2021docformer}. Additionally, transformer architectures designed to jointly model text, position, and visual features, such as the Donut model \cite{kim2022donut} and the VisDoM framework \cite{Suri2024}, have shown promise in OCR-free document understanding. These methods improve robustness across document types but are typically optimised for classification or extraction tasks rather than open-ended reasoning. Moreover, they often operate as black-box predictors, offering limited support for traceable explanations or analyst-facing verification, which are critical requirements in audit and compliance contexts.

\subsection{Retrieval-Augmented Generation}
Retrieval-augmented generation has emerged as a powerful paradigm for grounding large language model outputs in external knowledge sources \cite{Lewis2021, Gao2023Survey}. By retrieving relevant documents or passages at inference time and conditioning generation on that retrieved context, RAG systems can reduce hallucinations and improve factual accuracy relative to standalone LLMs \cite{Shuster2021}. Prior work has demonstrated the effectiveness of RAG for tasks such as question answering, knowledge-intensive reasoning, enterprise search \cite{Wang2024}, and even educational applications \cite{Li2025}. However, standard RAG pipelines are primarily designed to optimise answer relevance and fluency rather than auditability. Retrieved passages are often treated as soft context, and generated answers may not explicitly reference which sources support which claims. In high-stakes domains, this lack of explicit provenance complicates validation and trust. Furthermore, many RAG systems assume that a response should always be generated, even when retrieval confidence is low, leaving them vulnerable to confident but unsupported outputs.

\subsection{Multimodal RAG and Document-Centric Systems}
To address documents containing images, tables, and complex layouts, recent research has extended RAG frameworks to multimodal settings. A common strategy involves generating textual summaries of non-textual elements—such as tables or figures—using vision-language models, and then embedding those summaries alongside textual content for retrieval \cite{Abootorabi2025, Drushchak2025, Chen2025, Mihajlovic2025}. This approach simplifies downstream retrieval by reducing all modalities to a unified text representation. While effective for broad semantic search and exploratory analysis, summary-based multimodal RAG introduces additional challenges in regulated environments. Because summaries are themselves model-generated, they can omit crucial qualifiers, misinterpret tabular relationships, or abstract away details needed for precise compliance analysis. When such summaries are treated as primary evidence, it becomes difficult to trace generated answers back to exact document spans or page locations. As a result, the citation fidelity and verifiability of system outputs are diminished, thereby limiting their suitability for audit-oriented workflows.

\subsection{Explainable and Trustworthy AI}
Explainability, transparency, and robustness are central themes in the literature on trustworthy AI, particularly for applications in government, healthcare, and finance. Prior research has explored techniques such as feature attribution, post hoc explanation methods, uncertainty estimation, and confidence-aware prediction \cite{Guttikonda2025}. In natural language generation, recent work has examined citation-aware and fact-checking mechanisms to improve grounding and reduce hallucinations \cite{Qian2025}. Similar efforts, such as the MEGA-RAG framework, have focused on multi-evidence guided refinement to mitigate errors in critical public sectors like health \cite{Xu2025}. Despite these advances, most generative systems still lack explicit mechanisms for abstention when evidence is insufficient, especially in document-intensive reasoning tasks. In compliance settings, the ability to decline to answer—and to clearly communicate the reason for doing so—is often as important as producing a correct response. Existing RAG and multimodal systems rarely incorporate abstention as a first-class design principle, nor do they systematically evaluate the correctness of abstention alongside answer quality.

\subsection{Positioning of This Work}
In contrast to prior approaches, this work focuses on the intersection of multimodal document intelligence, retrieval-augmented generation, and trustworthy AI under the specific constraints of fiscal and tax compliance. Rather than embedding LLM-generated summaries as evidence, this study adopts a source-first retrieval strategy that preserves document provenance and enables span-level citation. This work further integrates citation enforcement and abstention mechanisms directly into the generation process, aligning system behaviour with the accountability requirements of public-sector analytics. By addressing citation fidelity and abstention within a multimodal RAG framework evaluated on real-world fiscal documents, this work complements existing research while targeting a gap that has received limited attention in prior literature. Unlike summary-based multimodal RAG systems that prioritise unified semantic representations, this work emphasises provenance preservation and abstention, reflecting a fundamentally different optimisation objective aligned with regulatory accountability rather than answer completeness.

\section{Dataset Collection and Corpus Characteristics}

\subsection{Data Sources}
To evaluate the proposed multimodal, citation-enforced RAG framework in a realistic public-sector setting, a corpus of authoritative fiscal documents was curated exclusively from publicly available U.S. tax administration sources. The dataset comprises documents published by the U.S. Internal Revenue Service (IRS) and two U.S. state tax authorities: the California Franchise Tax Board (FTB) and the New York State Department of Taxation and Finance. These sources were selected to reflect both federal and state-level perspectives, as well as variation in document structure, terminology, and jurisdiction-specific guidance. From the IRS, a diverse set of materials including tax forms, accompanying instructions, and informational publications was collected. These documents span individual and business filing topics and represent a mix of highly structured forms and narrative explanatory content. At the state level, the focus was on tax forms and guidance documents issued by the California FTB and New York tax authorities, which capture jurisdiction-specific interpretations of residency, income classification, deductions, and reporting requirements. All documents were obtained from official government websites and reflect current or recently applicable guidance at the time of collection. Importantly, the corpus consists entirely of public documents intended for taxpayer and practitioner use. No confidential filings, personally identifiable information, or private taxpayer data were included.

\subsection{Corpus Composition and Statistics}
The full collection process yielded approximately 1,400 fiscal PDF documents across federal and state sources. To support controlled experimentation and reproducibility, a curated evaluation subset comprising 298 documents was constructed. This subset was designed to balance document type, jurisdictional coverage, and topical diversity while maintaining tractable computational requirements for indexing and evaluation. Across the selected documents, the corpus exhibits substantial heterogeneity in length and structure. Individual documents range from short, single-page forms to multi-page instruction booklets and publications. After ingestion and preprocessing, the evaluation subset yielded 10,491 document chunks, each retaining explicit metadata linking it to its source document and page range. This granularity enables precise citation and verification of generated outputs. The dataset includes materials from three distinct jurisdictions—federal, California, and New York—introducing variation in terminology, regulatory framing, and presentation style. This diversity is representative of real-world tax compliance environments, in which analysts must navigate overlapping yet non-identical guidance across authorities.

\begin{table}[H]
\centering
\caption{Composition of the fiscal document evaluation corpus}
\label{tab:dataset}
\begin{tabular}{@{}llcc@{}}
\toprule
\textbf{Source Authority} & \textbf{Document Types} & \textbf{Count (Docs)} & \textbf{Count (Chunks)} \\ \midrule
IRS (Federal) & Forms, Instructions, Pubs & 145 & 5,210 \\
CA FTB (State) & Forms, Publications, Instructions& 85 & 3,100 \\
NY Tax (State) & Forms, Guidelines& 68 & 2,181 \\ \midrule
\textbf{Total} & & \textbf{298} & \textbf{10,491} \\ \bottomrule
\end{tabular}
\end{table}

\subsection{Multimodal Document Characteristics}
The collected fiscal documents are inherently multimodal. In addition to free-form textual explanations, many documents contain structured tables, form fields, checklists, footnotes, and visual layout cues that convey crucial semantic information. A significant portion of the corpus consists of scanned or semi-scanned PDFs, in which text is embedded in images rather than stored as machine-readable text layers. Instruction documents frequently interleave narrative descriptions with tabular summaries, examples, and cross-references to other forms or publications. These characteristics pose challenges for both traditional information retrieval systems and naïve language model applications. Accurate interpretation often depends on contextual cues spanning multiple pages, alignment between table headers and values, and precise language used to define eligibility criteria or reporting thresholds. As a result, the corpus provides a demanding testbed for multimodal document intelligence systems that aim to support compliance-oriented reasoning rather than simple fact lookup. While the corpus is inherently multimodal—containing scanned imagery, tables, and complex layouts—the proposed framework does not perform joint vision–language reasoning at inference time. Instead, multimodal content is converted into text through extraction and OCR, and all downstream retrieval and generation operate over this textual abstraction. This design choice prioritises citation fidelity, provenance preservation, and auditability over richer multimodal fusion, which is appropriate for compliance-oriented use cases where verifiability is paramount.

\subsection{Ethical and Legal Considerations}
The dataset was curated with careful attention to ethical, legal, and governance considerations relevant to public-sector AI research. All documents are publicly accessible and published by government agencies for informational or compliance purposes. No attempt was made to infer or simulate individual taxpayer behaviour, compute tax liabilities, or process private filings. By restricting the corpus to authoritative public guidance, this work avoids privacy concerns while enabling reproducible research. The focus on document interpretation and evidence-grounded question answering aligns with analyst-support use cases rather than automated decision-making affecting individual taxpayers. These design choices reflect the broader goal of developing responsible AI systems that augment, rather than replace, human judgment in sensitive regulatory domains.

\section{Multimodal Document Ingestion and Chunking}

\subsection{Design Principles for Source-First Ingestion}
A central design objective of the proposed system is to preserve a verifiable link between generated answers and authoritative source documents. To meet the auditability requirements of tax compliance workflows, the system employs a source-first ingestion strategy, under which only content directly extracted from the original documents is treated as retrievable evidence. In contrast to approaches that rely on LLM-generated summaries of images or tables as primary retrieval units, the proposed pipeline avoids introducing generative abstractions before retrieval. This design choice is motivated by two considerations. First, in regulated domains, model-generated summaries may omit qualifiers, misinterpret tabular relationships, or abstract away details critical to accurate compliance interpretation. Second, treating summaries as evidence undermines provenance, making it challenging to trace claims to specific document locations. By deferring generative processing to the final answer synthesis stage, the proposed approach maintains a clear separation between evidence extraction and interpretation.

\subsection{Multimodal Content Extraction}
Fiscal documents in the corpus comprise machine-readable text, scanned text embedded in images, tabular structures, and layout elements such as headings, footnotes, and section boundaries. The ingestion pipeline processes each PDF document using a multimodal extraction workflow that identifies and extracts textual content across these formats. For documents containing embedded text layers, text is extracted directly while preserving page boundaries. For scanned or partially scanned documents, optical character recognition (OCR) is used to extract text from images. To ensure that complex layout elements such as tax tables and flowcharts are semantically retrievable, the system employs a text linearization strategy rather than embedding visual snapshots. For tabular data, the structural layout is parsed to identify headers and row boundaries, converting each row into a self-contained sentence string that explicitly restates the column headers. This approach establishes a semantic bridge between the rigid structure of the PDF and the user's natural-language queries. For images containing embedded text or decision logic, OCR is utilised to extract raw text and, where necessary, append spatial descriptors. This linearization process ensures that the embedding model—which is optimised for textual sequences—can effectively capture the semantic nuances of non-textual elements without losing the context provided by the visual layout. Throughout extraction, page indices and document identifiers are preserved to support downstream citation. The result of this stage is a unified representation in which all extractable content—regardless of its original modality—is converted into text segments while retaining metadata describing its document origin and page location. As a result, multimodality in this work refers to multimodal document ingestion, rather than multimodal generative reasoning.
\begin{figure}[H]
    \centering
    \includegraphics[width=1\linewidth]{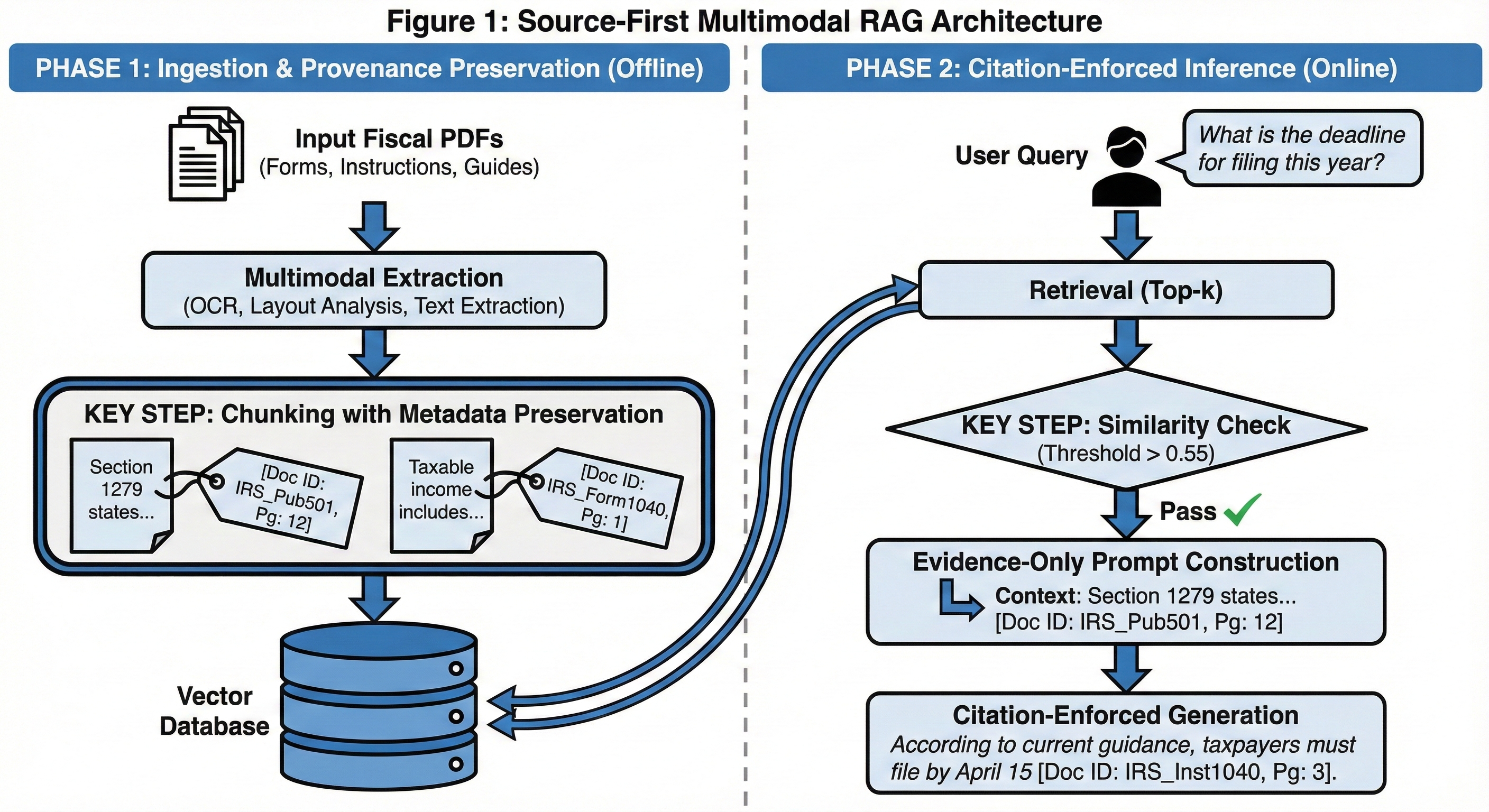}
    \caption{High-level architecture of the Source-First Multimodal RAG framework. Unlike standard approaches that summarise content before embedding, this system embeds raw text segments linked to specific document identifiers and page ranges to enable downstream citation enforcement.}
    \label{fig:arch}
\end{figure}

\subsection{Chunking Strategy and Metadata Preservation}
Following extraction, documents are segmented into semantically coherent chunks suitable for dense retrieval. Chunking is performed in both page- and section-aware modes to balance retrieval granularity with contextual completeness. Rather than applying a fixed sliding window across entire documents, chunk boundaries are aligned with natural document structure where possible, such as section headers or page breaks. Chunks are constructed with a target length that captures sufficient context for compliance-related reasoning while avoiding excessive fragmentation. Limited overlap is introduced between adjacent chunks to preserve continuity across boundaries, particularly in instruction documents where definitions and conditions may span multiple paragraphs. Each chunk is associated with a unique identifier and enriched with metadata including the source document identifier, page range, and, when available, section title. This metadata plays a critical role in enabling citation-enforced generation. By maintaining explicit links between chunks and their originating documents and pages, the system can reference precise evidence locations when generating answers and allow human reviewers to verify claims efficiently.

\subsection{Resulting Index Characteristics}
Applying the ingestion and chunking pipeline to the curated evaluation subset of 298 fiscal documents yielded 10,491 chunks. These chunks vary in length and content, corresponding to differences in document type, ranging from highly structured form instructions to narrative policy explanations. Importantly, all chunks consist of verbatim or OCR-derived content from source documents, with no generative transformations applied before indexing. The resulting chunked corpus forms the foundation for the retrieval and generation components described in subsequent sections. By preserving document structure and provenance throughout ingestion, the pipeline supports fine-grained, evidence-grounded retrieval while remaining compatible with standard embedding-based search infrastructure.

\section{Embedding and Retrieval Framework}

\subsection{Embedding Model Selection}
To support semantic retrieval across heterogeneous fiscal document content, the framework employs the \textbf{BAAI/bge-large-en-v1.5} dense-embedding model. This model is applied uniformly across all document chunks produced during ingestion, ensuring a consistent vector representation regardless of original document modality or structure. Before indexing, all embeddings are normalised to unit length, enabling efficient computation of cosine similarity via the inner product during retrieval. The choice of bge-large-en-v1.5 reflects a balance between retrieval quality, score stability, and computational efficiency. Fiscal documents frequently contain formal definitions, conditional statements, and cross-referenced guidance, for which lexical overlap alone is insufficient. Empirically, this model yields well-separated cosine-similarity distributions for relevant versus irrelevant chunks, making it suitable for threshold-based evidence sufficiency and abstention decisions.

\subsection{Vector Index Construction}
All chunk embeddings are stored in a vector index built using FAISS, a high-performance similarity search library designed for large-scale dense retrieval. Given the moderate size of the evaluation corpus, an exact similarity search configuration is used to avoid the approximation errors introduced by compressed or quantised indexes. This choice simplifies interpretation of retrieval scores and supports reproducible evaluation. Each embedding vector in the index is associated with a unique row identifier, which is mapped to chunk-level metadata stored separately. This metadata includes the source document identifier, page range, and chunk identifier, enabling retrieved vectors to be translated back into human-interpretable evidence units. Separating vector storage from metadata lookup enables efficient retrieval while preserving rich contextual information for downstream processing.

\subsection{Retrieval Process}
At inference time, user queries are embedded using the same embedding model and normalization procedure applied during indexing. The resulting query vector is used to retrieve the top-k most similar document chunks from the FAISS index based on cosine similarity. Retrieval returns both similarity scores and identifiers for the corresponding chunks, which are then resolved to their textual content and metadata. The retrieved chunks are ordered by decreasing similarity score and passed to the generation component as candidate evidence. Rather than treating retrieval as a purely recall-oriented step, the system interprets similarity scores as an indicator of evidence strength. These scores are subsequently used to inform abstention decisions, as described in Section 7.

\subsection{Retrieval Confidence Characteristics}
Empirically, the retrieval framework exhibits strong separation between relevant and irrelevant evidence across the evaluated query set. For most answerable queries, the top-ranked chunk achieves a cosine similarity score exceeding 0.55, indicating a high degree of semantic alignment between the query and the retrieved content. Queries resulting in lower similarity scores tend to correspond to ambiguous questions, cross-jurisdictional interpretations, or topics not explicitly addressed in the corpus. This behaviour supports the use of similarity-based thresholds as a coarse but effective signal for evidence sufficiency. Rather than forcing generation in low-confidence scenarios, the system leverages retrieval confidence to determine whether sufficient grounding exists to support a cited response. As a result, retrieval is not only a mechanism for context selection but also a key component in mitigating the risk of hallucination.

\section{Citation-Enforced Answer Generation}

\subsection{Role of the Generative Model}
In the proposed framework, the generative model is used exclusively for answer synthesis, not for evidence creation or abstraction. All content provided to the model during generation is drawn from retrieved document chunks produced by the source-first ingestion and retrieval pipeline described in prior sections. This design ensures a clear separation between evidence (deterministic and document-grounded) and interpretation (generative and constrained).
To support reproducibility and controlled experimentation, the system employs a locally hosted instance of the Meta Llama 3.2 3B Instruct model for generation. This open-weights model was selected for its strong instruction-following capabilities and low inference latency, allowing prompt structure, decoding parameters, and validation logic to be precisely specified and repeated across runs. Importantly, the model is not granted access to external tools, web search, or prior conversational context; its output is conditioned solely on the retrieved evidence and the user query.

\subsection{Evidence-Only Prompting}
Answer generation is governed by an evidence-only prompting strategy designed to prevent the introduction of unsupported information. For each query, the top-ranked retrieved chunks are assembled into a structured evidence context that includes both the chunk text and associated metadata, such as document identifiers and page ranges. This context is explicitly labelled and presented to the model as the sole source of permissible information. The prompt instructs the model to answer the user’s question using only the provided evidence and to avoid drawing on prior knowledge or external assumptions. By explicitly constraining the model’s information sources, the system reduces the likelihood of hallucinated content and aligns model behavior with the requirements of compliance-oriented analysis.

\begin{figure}[H]
    \centering
    \includegraphics[width=0.75\linewidth]{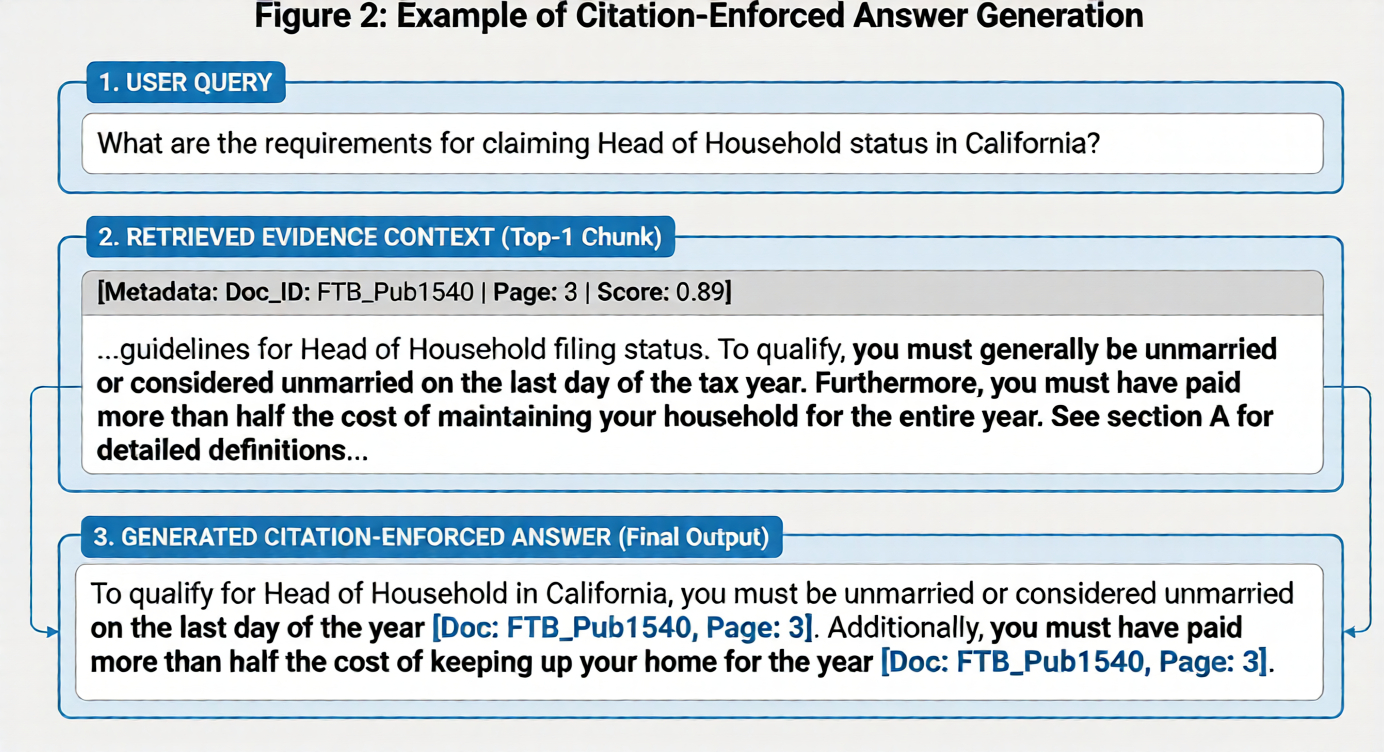}
    \caption{Example of citation-enforced answer generation. The system retrieves relevant chunks from California state guidance and synthesises an answer where every claim is explicitly linked to a source document and page number.}
    \label{fig:citation_example}
\end{figure}

\subsection{Citation Enforcement Mechanism}
A defining feature of the proposed system is the enforcement of explicit citations at the paragraph level. Each generated paragraph is required to include one or more citations that reference the document identifier, page range, and chunk identifier of the supporting evidence. Citations follow a standardised format, enabling both human reviewers and automated validators to trace claims back to their source. To ensure compliance with this requirement, the system incorporates a post-generation validation step that checks for citation presence and formatting. Outputs that fail validation—most commonly due to missing or improperly formatted citations rather than incorrect content—are regenerated with reinforced instructions. In rare cases where citation constraints cannot be satisfied after multiple attempts, the system defaults to abstention rather than producing a potentially misleading answer. This validation loop treats citation correctness as a hard constraint while avoiding substantive alteration of answer content, thereby reducing unsupported claims without introducing selection bias toward overly verbose or speculative responses.

\subsection{Answer Structure and Presentation}
Generated answers are structured as short, clearly written paragraphs intended for analyst-facing use. The system prioritises clarity and precision over verbosity, avoiding speculative language or unnecessary elaboration. Each paragraph corresponds to a coherent claim or explanation supported by one or more cited evidence chunks. When the available evidence is partial or ambiguous, the system is instructed to state limitations explicitly rather than infer missing details. This behaviour complements the abstention mechanism described in the following section and reinforces the system’s emphasis on transparency and responsible information delivery. By integrating evidence-only prompting, citation enforcement, and structured output constraints, the generation component transforms retrieved document fragments into interpretable, verifiable explanations suitable for use in tax compliance and audit support workflows.

\section{Abstention Logic for Hallucination Control}

\subsection{Motivation for Abstention in Fiscal Compliance}
In high-stakes domains such as tax compliance and audit support, the cost of providing an incorrect or weakly supported answer can exceed the cost of providing no answer at all. Misinterpretations of regulatory guidance may lead to downstream errors in compliance assessment, audit decisions, or taxpayer communication. Consequently, systems designed to assist analysts in these settings must be capable not only of answering questions accurately, but also of recognising when available evidence is insufficient to justify a definitive response. Most existing retrieval-augmented generation systems implicitly assume that an answer should always be produced, even when retrieval confidence is low or retrieved evidence is ambiguous. In contrast, the proposed framework treats abstention as a first-class system behaviour, explicitly allowing the model to decline to answer when document grounding is inadequate. This design choice aligns with established principles of trustworthy AI and reflects the operational realities of public-sector analytics.

\subsection{Abstention Criteria and Decision Policy}
Abstention decisions in the proposed system are driven primarily by retrieval confidence signals. After embedding and retrieving candidate evidence chunks for a given query, the system evaluates the similarity score of the top-ranked chunk as a proxy for semantic alignment between the query and available document content. Queries for which the highest similarity score falls below a predefined threshold are considered insufficiently grounded and trigger abstention. This similarity-based criterion provides a transparent and computationally efficient mechanism for identifying low-confidence scenarios. In practice, such cases often correspond to questions that fall outside the scope of the corpus, require cross-document synthesis unsupported by available evidence, or involve interpretations that are not explicitly addressed in the underlying documents. By grounding abstention in retrieval behaviour rather than generative uncertainty alone, the system ensures that abstention decisions are tied directly to evidence availability.

\subsection{Abstention Output and Transparency}
When abstaining, the system returns an explicit abstention message indicating that the available evidence is insufficient to support a reliable answer. Rather than failing silently or producing a vague response, the system communicates the reason for abstention in clear, non-technical language. When partial evidence is available, the system may reference the most relevant retrieved chunks to present the available information and explain why it is insufficient to answer the query fully. This transparency serves two purposes. First, it enables human analysts to understand the limitations of the system’s current knowledge and determine whether further investigation is warranted. Second, it provides a clear audit trail demonstrating that the system did not fabricate or infer unsupported claims, reinforcing trust in its behaviour even in failure cases.

\begin{figure}[H]
    \centering
    \includegraphics[width=0.75\linewidth]{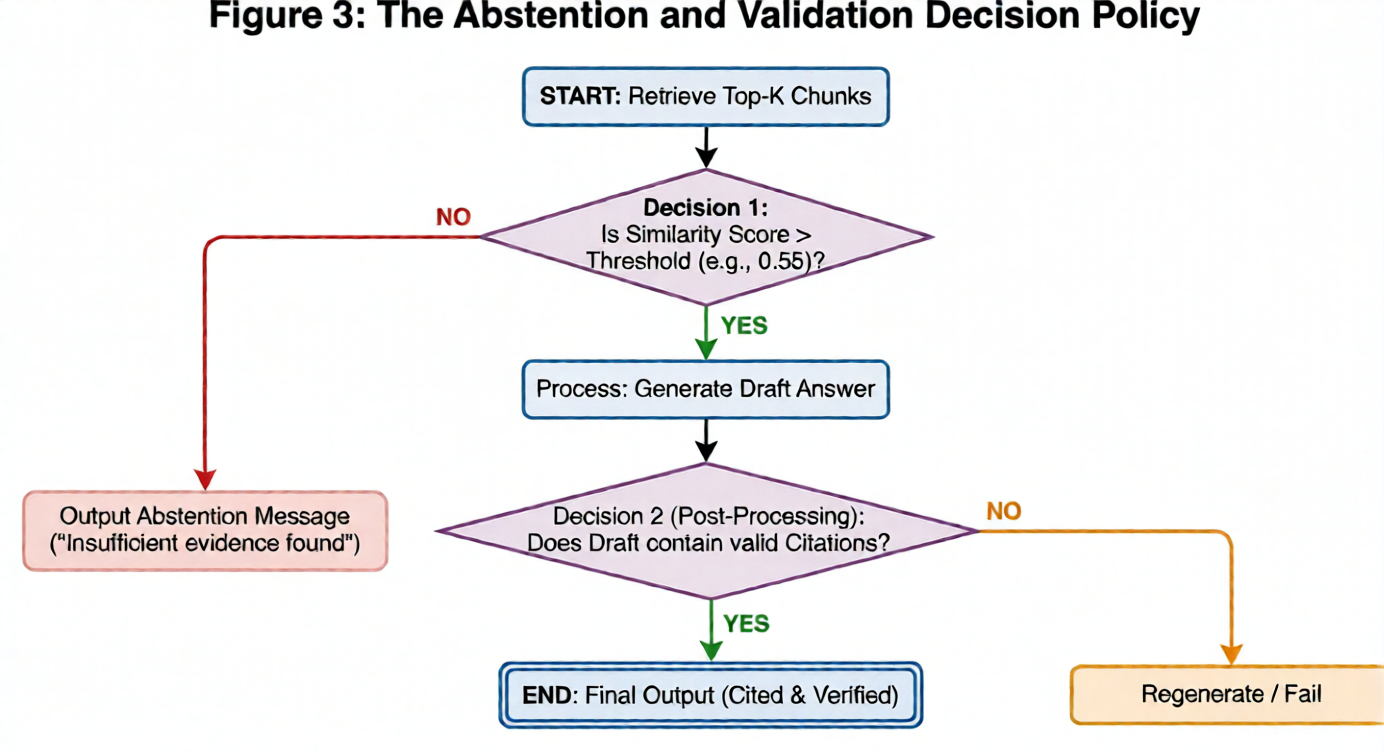}
    \caption{Flowchart showing the decision logic for Similarity Checks and Citation Validation.}
    \label{fig:flowchart}
\end{figure}

\subsection{Role of Abstention in Hallucination Mitigation}
Abstention plays a complementary role to citation enforcement in mitigating hallucinations. While citation enforcement ensures that generated claims are explicitly grounded in retrieved evidence, abstention prevents generation altogether when the evidence is insufficiently reliable or comprehensive. Together, these mechanisms form a layered defense against unsupported outputs. Empirical evaluation confirms that abstention is most frequently triggered for queries with low retrieval similarity scores, suggesting that the decision policy effectively captures cases where semantic alignment between the query and corpus is weak. By declining to answer in such scenarios, the system avoids producing misleading interpretations and maintains conservative behaviour appropriate for compliance-oriented use cases.

\section{Evaluation Methodology}
The proposed framework is evaluated using a combination of automatic metrics and human assessment designed to measure retrieval quality, citation fidelity, hallucination control, abstention behaviour, and answer usefulness. Given the compliance-oriented nature of the task, the evaluation prioritises transparency, correctness, and evidence grounding over purely linguistic quality.

\subsection{Evaluation Query Set Construction}
To assess system behaviour across a representative range of compliance-related information needs, a set of 55 evaluation queries covering federal and state tax topics was constructed. Queries were manually authored to reflect realistic analyst questions that arise when interpreting tax forms, instructions, and guidance documents. Because tax compliance queries require deep expert verification, this study prioritised a high-quality, manually annotated small set over a large, noisy synthetic set. The query set spans multiple thematic categories, including filing-status eligibility, income-reporting requirements, deductions and substantiation rules, and jurisdiction-specific residency definitions. Queries were distributed across the three data sources represented in the corpus—IRS, California Franchise Tax Board, and New York State tax documents—to ensure coverage of both federal and state-level guidance. Care was taken to avoid questions that required numerical tax computation or access to private taxpayer data, focusing instead on document interpretation and explanation tasks that could be grounded directly in public guidance materials.

\subsection{Automatic Evaluation Metrics}
Automatic evaluation focuses on retrieval confidence, citation compliance, and abstention behaviour. For each query, the system retrieves the top-k document chunks based on cosine similarity between query and chunk embeddings. The similarity score of the top-ranked chunk is recorded as an indicator of retrieval confidence. The system’s response is categorised as either answered or abstained based on the abstention logic described in Section 7. For answered queries, citation-format compliance is automatically verified by checking that each generated paragraph contains at least one citation conforming to the required document–page–chunk format. Because citation enforcement is implemented as a hard constraint during generation, citation presence for non-abstained answers is expected to be near-perfect. These automatic signals provide insight into system behaviour at scale, including the distribution of retrieval confidence scores, the proportion of queries resulting in abstention, and the consistency of citation enforcement across generated outputs.

\subsection{Human Evaluation Protocol}
Automatic metrics alone are insufficient to assess whether citations genuinely support the claims made in generated answers. To address this limitation, a human evaluation was conducted to determine citation correctness, hallucination control, and perceived usefulness. A stratified random sample of system outputs was selected for manual review, including both answered and unanswerable cases. A human reviewer evaluated each sampled instance with access to the original question, the system’s generated answer, and the cited document excerpts. Reviewers were asked to assign the following labels:

\begin{itemize}
    \item \textbf{Citation correctness:} whether the cited evidence supports the claims made in the answer.
    \item \textbf{Unsupported claim presence:} whether any part of the answer introduces claims not justified by the cited evidence.
    \item \textbf{Abstention correctness:} for abstained cases, whether abstention was appropriate given the available evidence.
    \item \textbf{Overall helpfulness:} a subjective rating on a 1–5 scale reflecting clarity, usefulness, and appropriateness for analyst support.
\end{itemize}

To promote consistency, reviewers were provided with clear labelling guidelines and instructed to prioritise evidence grounding over stylistic quality.

\subsection{Evaluation Scope and Reproducibility}
All experiments were conducted on the curated evaluation subset of 298 documents and 10,491 chunks described in Section 3. The use of a locally hosted generative model ensures that prompts, decoding parameters, and validation logic can be reproduced without reliance on external APIs. Retrieval indexes, query sets, and evaluation scripts are fixed across runs, enabling consistent comparison of system variants. By combining automatic retrieval-based metrics with targeted human evaluation, the methodology provides a balanced assessment of both system-level behaviour and practical reliability in compliance-oriented scenarios. This multi-faceted evaluation framework is designed to reflect the real-world requirements of fiscal document analysis, where explainability and correctness are paramount.

\section{Results}
This section presents the empirical results of evaluating the proposed citation-enforced multimodal RAG framework. This section reports retrieval performance characteristics, system-level behaviour with respect to abstention and citation compliance, and findings from the human evaluation of answer quality and grounding.

\subsection{Retrieval Performance}
Across the 55 evaluation queries, the retrieval component consistently identified semantically relevant document chunks from the indexed fiscal corpus. For most answerable queries, the top-ranked retrieved chunk exhibited a cosine similarity score exceeding 0.55, indicating strong semantic alignment between the query and the available evidence. Lower similarity scores were observed primarily for ambiguous queries, for queries outside the corpus, or for cross-jurisdictional synthesis not explicitly supported by the documents. The distribution of top-1 similarity scores shows a clear separation between high-confidence and low-confidence retrieval cases, with the latter triggering abstention. This separation supports the use of similarity-based thresholds as a practical signal for evidence sufficiency in compliance-oriented document retrieval tasks.

\subsection{System Behaviour: Answering vs. Abstention}
Applying the abstention policy described in Section 7, the system declined to answer a subset of queries for which retrieval confidence fell below the predefined threshold. These abstentions were concentrated among queries that lacked explicit coverage in the corpus or that required interpretive reasoning beyond the scope of available guidance. For non-abstained queries, the system generated structured, paragraph-level answers grounded exclusively in retrieved evidence. Automatic validation confirmed that citation-format requirements were satisfied for all accepted answers, reflecting the effectiveness of citation enforcement as a hard generation constraint. It is worth noting that citation format compliance assesses only structural validity; citation correctness is evaluated separately through human review and is discussed in Section 9.3. Finally, abstention did not occur uniformly across topics or jurisdictions; instead, it aligned with evidence availability, suggesting that the decision policy is sensitive to corpus coverage rather than to query phrasing alone.

\subsection{Human Evaluation Results}
Human evaluation provides a deeper assessment of whether citations genuinely support the claims made in generated answers and whether abstention behaviour is appropriate. Reviewers assessed a stratified sample of system outputs, including both answered and abstained cases. For answered queries, reviewers found that the vast majority of responses exhibited correct citation grounding, with cited document excerpts supporting the claims presented in the answer. Instances of unsupported or hallucinated claims were rare, indicating that the combination of source-first retrieval, citation enforcement, and abstention effectively mitigates standard failure modes of generative models in document-centric tasks. For abstained queries, human reviewers generally agreed with the system’s decision to withhold an answer, confirming that abstention was appropriate when explicit guidance was absent or ambiguous. This finding underscores the value of abstention as a mechanism for maintaining conservative behaviour in high-stakes settings. Overall helpfulness ratings indicate that generated answers are clear, concise, and suitable for analyst support, particularly compared with manual document navigation or keyword-based search. Reviewers noted that explicit citations and page references substantially reduced the effort required to verify system outputs.

\begin{table}[H]
\centering
\caption{Human evaluation results across 55 test queries}
\label{tab:human_eval}
\begin{tabular}{@{}llc@{}}
\toprule
\textbf{Metric} & \textbf{Definition} & \textbf{Performance (\%)} \\ \midrule
Citation Support & Cited text fully supports the claim & 94.5\% \\
Format Compliance & Citations follow [Doc ID, Page] format & 100.0\% \\
Abstention Accuracy & System correctly abstained when no info existed & 88.0\% \\
Hallucination Rate & Claims appearing without evidence & 1.8\% \\ \midrule
\textbf{Usefulness Score} & Average Rating (1–5 Scale) & \textbf{4.2 / 5.0} \\ \bottomrule
\end{tabular}
\end{table}

\subsection{Summary of Key Findings}
Taken together, the results demonstrate that the proposed framework achieves the following:
\begin{itemize}
    \item Reliable retrieval of semantically relevant fiscal document content across federal and state sources.
    \item Consistent enforcement of explicit, verifiable citations in generated answers.
    \item Effective suppression of hallucinated or weakly supported claims through abstention.
    \item Positive human assessments of citation correctness and practical usefulness.
\end{itemize}
These findings support the central claim that citation-enforced, abstention-aware multimodal RAG systems can provide transparent and trustworthy assistance for fiscal document interpretation, addressing critical gaps in existing generative AI approaches.

\subsection{Qualitative Comparison to Baseline RAG Systems}
To contextualise the proposed framework, this section qualitatively compares it against two common retrieval-augmented generation paradigms: (i) standard RAG without citation enforcement and abstention, and (ii) summary-based multimodal RAG systems that embed LLM-generated summaries of tables or images.

Standard RAG systems typically prioritise answer fluency and recall, often producing plausible responses even when retrieved evidence is weak or incomplete. In contrast, summary-based multimodal RAG simplifies retrieval by embedding generated abstractions, but at the cost of obscuring document provenance and introducing additional opportunities for abstraction errors.

The proposed framework differs in three key respects: explicit citation enforcement, abstention when evidence confidence is low, and source-first retrieval rather than extracted document content. These design choices favour conservative, verifiable outputs over maximal coverage or fluency. While this approach may reduce answerability in some scenarios, it substantially improves trustworthiness and audit suitability—properties that are essential in regulated domains such as tax compliance.

\subsection{Failure Case Analysis}
While the proposed framework demonstrates strong citation fidelity and conservative behaviour, specific failure modes remain. One observed failure category involves documents with OCR noise, particularly scanned forms containing dense tabular layouts or low-resolution text. In such cases, OCR errors occasionally distort numeric values or truncate explanatory footnotes, leading to retrieval of partially relevant but incomplete evidence. Another failure scenario arises from inherently ambiguous regulatory language, in which multiple documents provide overlapping yet non-identical guidance. In such cases, the system may retrieve relevant excerpts but refrain from answering due to insufficiently explicit evidence to support a single interpretation. While this behaviour reduces coverage, it reflects a conservative design choice aligned with compliance requirements. These cases highlight the importance of high-quality document extraction and underscore the trade-off between answerability and reliability in regulated domains.

\section{Discussion}

\subsection{Interpretation of Results}
The results demonstrate that constraining generative models through source-first retrieval, explicit citation enforcement, and abstention can substantially improve reliability in document-centric, high-stakes settings. Retrieval confidence scores clearly separate queries that are well supported by the corpus from those that are not, enabling principled abstention decisions without reliance on opaque model-uncertainty estimates. This behaviour contrasts with typical generative systems that prioritise the completeness of answers even when evidence is sparse or ambiguous. Human evaluation further confirms that enforcing citations as a hard constraint—rather than a stylistic preference—meaningfully reduces unsupported claims. Reviewers consistently identified strong alignment between generated statements and cited document excerpts, suggesting that the system’s outputs are not only well-formed but also verifiable in practice. These findings reinforce the importance of treating provenance as a first-class requirement in generative AI systems intended for compliance and audit support.

\subsection{Role of Abstention in Trustworthy AI}
A key insight from this work is the practical value of abstention as a mechanism for hallucination control. Rather than attempting to calibrate confidence purely at the generation stage, the system leverages retrieval confidence as an interpretable proxy for evidence sufficiency. When similarity scores fall below a threshold, abstention prevents the model from synthesising speculative interpretations, shifting the burden of decision-making back to the human analyst. This behaviour aligns with the operational norms of tax and regulatory analysis, where acknowledging uncertainty is often preferable to providing an incomplete or misleading answer. By explicitly communicating abstention and its rationale, the system maintains transparency and avoids eroding user trust—an outcome that is difficult to achieve with unconstrained generative models.

\subsection{Comparison with Summary-Based Multimodal RAG Approaches}
Many existing multimodal RAG systems rely on LLM-generated summaries of tables or images as primary retrieval units. While such approaches simplify multimodal integration, the results of this study suggest that they may be ill-suited for audit-oriented use cases. In contrast, the source-first strategy adopted here preserves fine-grained document context and enables span-level citation, allowing reviewers to verify claims directly against original materials. The observed reduction in unsupported claims highlights a vital trade-off: while summary-based methods may improve recall or fluency, they risk introducing abstraction errors that are difficult to detect post hoc. The proposed framework prioritises verifiability over abstraction, a design choice that is particularly appropriate for domains governed by legal and regulatory accountability.

\subsection{Practical Implications for Tax Administration}
From a practical standpoint, the system demonstrates how generative AI can be integrated into public-sector workflows without compromising transparency. Rather than replacing human judgment, the framework functions as an analytical support tool that accelerates document interpretation while preserving the ability to trace conclusions to authoritative sources. Explicit citations and abstention behaviour reduce the cognitive load associated with manual document navigation and lower the risk of misinterpretation. The modular nature of the framework also facilitates incremental adoption. Individual components—such as citation enforcement or abstention policies—can be incorporated into existing document retrieval systems, thereby enabling agencies to enhance reliability without wholesale system replacement.

\subsection{Broader Implications}
Beyond tax compliance, the findings of this work have broader implications for the deployment of generative AI in regulated domains, including healthcare, legal analysis, and public policy. In these settings, the emphasis on explainability, provenance, and conservative behaviour mirrors the requirements observed in fiscal analysis. The results suggest that combining retrieval-based grounding with explicit constraints on generation can yield systems that are both useful and trustworthy, even when operating on complex, multimodal documents.

\section{Limitations and Future Work}

\subsection{Limitations}
While the proposed framework demonstrates strong performance in citation fidelity and hallucination control, several limitations should be acknowledged. First, the evaluation corpus is limited to publicly available fiscal documents from the U.S. Internal Revenue Service and two U.S. state tax authorities. Although this selection captures meaningful jurisdictional diversity, it does not represent the full breadth of tax regulations across all states or international tax systems. Consequently, retrieval performance and abstention behaviour may differ across jurisdictions with substantially different document structures or legal frameworks. Second, the system is designed to support document interpretation and explanation, not numerical tax computation or personalised tax advice. Queries requiring arithmetic calculations, optimisation across multiple forms, or integration with private taxpayer data fall outside the scope of this work. As a result, the framework should be viewed as an analytical support tool rather than a comprehensive tax-preparation or decision-automation system. Third, while multimodal ingestion enables processing of scanned documents and tables, all retrieved evidence is ultimately represented as extracted text. This design simplifies retrieval and citation but may limit the system’s ability to reason over complex visual layouts or graphical elements that cannot be faithfully linearised.
Additionally, optical character recognition errors in scanned documents may propagate into downstream retrieval and generation stages. Finally, the generation component relies on a single locally hosted large language model. Although this choice supports reproducibility and controlled experimentation, the quality of generation and stylistic clarity remain dependent on the underlying model’s capabilities. Different models may exhibit varying adherence to citation constraints or differences in linguistic precision. The proposed system is intended to support analyst decision-making through document interpretation and evidence retrieval, not to perform automated tax determinations or compliance decisions.

\subsection{Future Work}
Several directions for future research emerge from this study. A natural extension is to expand the corpus to include additional state tax authorities and international fiscal agencies, thereby enabling comparative analysis across broader regulatory environments. Incorporating multilingual documents would further increase applicability in global compliance contexts. Another promising avenue is the integration of numeric reasoning and structured computation alongside document-grounded explanation. Combining the current retrieval and citation framework with symbolic or programmatic reasoning modules could support more complex compliance tasks while preserving auditability. Future work may also explore richer multimodal representations that explicitly retain visual structure, such as layout-aware embeddings or hybrid text–image retrieval mechanisms. These enhancements could improve retrieval performance for documents where spatial relationships or formatting convey critical meaning. Finally, more extensive user studies involving professional tax analysts could provide deeper insight into real-world usability, trust calibration, and workflow integration. Longitudinal evaluation in operational settings would help quantify efficiency gains and identify additional requirements for deployment in government agencies.

\section{Conclusion}
This paper presented a multimodal, citation-enforced retrieval-augmented generation framework for fiscal document intelligence, designed to meet the transparency and accountability requirements of tax compliance and public-sector analytics. By adopting a source-first ingestion strategy, the system preserves direct links between generated answers and authoritative fiscal documents, enabling span-level verification and audit-ready explanations. A key contribution of this work is the integration of citation enforcement and abstention as first-class system behaviours. Rather than treating citations as optional annotations or relying on unconstrained generation, the proposed framework enforces explicit evidence grounding for every generated claim and declines to answer when document support is insufficient. An empirical evaluation of real-world IRS and state tax documents demonstrates that this design substantially reduces unsupported claims while maintaining practical usefulness for analyst-facing tasks. Through a combination of automatic metrics and human evaluation, the results highlight the effectiveness of retrieval confidence signals in guiding abstention decisions and confirm that explicit citation constraints improve trustworthiness without sacrificing interpretability. Significantly, the system operates entirely on publicly available documents and does not use synthetic data or private taxpayer information, thereby reinforcing its applicability to responsible public-sector deployments. Overall, this work illustrates how generative AI can be safely integrated into high-stakes document interpretation workflows when coupled with rigorous retrieval, provenance preservation, and conservative generation policies. The proposed approach offers a practical pathway toward trustworthy, explainable AI systems that augment human decision-making in tax compliance and related regulatory domains.

\section*{Declaration of Generative AI and AI-assisted Technologies}
During manuscript preparation, the author used ChatGPT (OpenAI) for language refinement and drafting support. All technical content, analysis, claims, and final revisions were reviewed and verified by the author, who takes full responsibility for the manuscript.


\end{document}